\documentclass[conference]{IEEEtran}

\usepackage{cite}
\usepackage{graphicx}
\usepackage{amsmath}
\usepackage{bbm}
\usepackage{algorithmicx}
\usepackage{algpseudocode}
\usepackage{algorithm}
\usepackage{dsfont}
\usepackage{color}
\usepackage{xcolor}
\usepackage{pdfcomment}
\usepackage[normalem]{ulem}
\usepackage{pgfplots}
\pgfplotsset{compat=newest}
\usepackage{tikz}
\usetikzlibrary{positioning,shapes,shapes.geometric,calc,chains,shapes.multipart,arrows}

\usepackage[caption=false,font=footnotesize]{subfig}

\tikzset{
	rubberduck/.style={
		shape=isosceles triangle,
		fill=red!30,
		minimum height=18pt,
		minimum width=12pt,
		shape border rotate=#1,
		isosceles triangle stretches,
		inner sep=0pt,
	},
	ap/.style={rubberduck=+90},
	ducky/.style={rubberduck=-90},
	queue/.style={
		rectangle split,
		minimum width=2em,
		rectangle split parts=5,
		draw,
		anchor=south,
	}
}

\makeatletter
\newenvironment{varsubequations}[1]
 {%
  \addtocounter{equation}{-1}%
  \begin{subequations}
  \renewcommand{\theparentequation}{#1}%
  \def\@currentlabel{#1}%
 }
 {%
  \end{subequations}
 }
\makeatother

\newtheorem{theorem}{Theorem}

\newtheorem{proposition}{Proposition}

\DeclareMathOperator*{\maximize}{maximize}
\DeclareMathOperator{\subjectto}{subject~to}

\algnewcommand\algorithmicinput{\textbf{INPUT:}}
\algnewcommand\INPUT{\item[\algorithmicinput]}

\algnewcommand\algorithmicoutput{\textbf{OUTPUT:}}
\algnewcommand\OUTPUT{\item[\algorithmicoutput]}

\algnewcommand\algorithmicinit{\textbf{Initialization:}}
\algnewcommand\Init{\item[\algorithmicinit]}

\algnewcommand\algorithmicpost{\textbf{Post Processing:}}
\algnewcommand\Post{\item[\algorithmicpost]}

\newcommand{\EW}[1]{#1}
\renewcommand{\sout}[1]{}

\newcommand{\br}{\boldsymbol{r}}

\newcommand{\bw}{\boldsymbol{w}}
\newcommand{\bx}{\boldsymbol{x}}
\newcommand{\by}{\boldsymbol{y}}
\newcommand{\bz}{\boldsymbol{z}}
\newcommand{\bh}{\boldsymbol{h}}

\newcommand{\NN}{\mathcal{N}}
\newcommand{\KK}{\mathcal{K}}

\newcommand{\Kcal}{\mathcal{K}}

\begin{document}
\title{Scalable Spectrum Allocation for Large Networks Based on Sparse Optimization\footnote{This
work was supported by a grant from Futurewei.}}

\author{\IEEEauthorblockN{Binnan Zhuang}
\IEEEauthorblockA{Modem R\&D Lab
\\Samsung Semiconductor, Inc.
\\San Diego, CA}
\and
\IEEEauthorblockN{Dongning Guo, Ermin Wei, and Michael L. Honig}
\IEEEauthorblockA{Department of Electrical Engineering and Computer Science
\\Northwestern University
\\Evanston, IL}}

\maketitle

\begin{abstract}
Joint allocation of spectrum and user association is considered for a large cellular network. The objective is to optimize a network utility function such as average delay given traffic statistics collected over a slow timescale. A key challenge is scalability: given $n$ Access Points (APs),
there are $O(2^n)$ ways in which the APs can
share the spectrum. The number of variables is reduced from $O(2^n)$ to $O(nk)$, where $k$ is the number of users, by optimizing over local overlapping neighborhoods,
defined by interference conditions, and by exploiting the existence of
sparse solutions in which the spectrum is divided into $k+1$ segments.
We reformulate the problem by optimizing the assignment of subsets
of active APs to those segments.
An $\ell_0$ constraint enforces a one-to-one mapping of subsets to spectrum,
and an iterative (reweighted $\ell_1$) algorithm is used to find an approximate solution.
Numerical results for a network with 100 APs
serving several hundred users show the proposed method achieves
a substantial increase in total throughput relative to benchmark schemes.

\end{abstract}

\section{Introduction}
\label{sec:Intro}

Heterogeneous cellular networks with dense deployment of access points (APs)
are anticipated to be a major component of 5G networks.
A challenge with such dense deployments is interference management.
Coordinated radio resource allocation across multiple cells
is one approach for mitigating inter-cell interference.
That includes
joint scheduling across multiple cells
over a fast timescale~\cite{yu2013multicell,wang2013joint,HonLuo13TransSP}
as well as the assignment of resources across cells
over a slower timescale~\cite{zhuang2014traffic,kuang2014joint}.
Whereas the former approach requires instantaneous knowledge of channel gains,
the latter approach relies on statistical knowledge of interference.

We consider the joint allocation of spectrum and user association in a large network
with many APs. The objective is to optimize a network utility, such as average delay,
given traffic statistics over a geographic region that change slowly relative
to channel fading. Our approach builds on our prior work
\cite{zhuang2016scalable,zhuang2014traffic} in which for a network of $n$ APs
and $k$ mobiles (or User Equipments (UEs)), the spectrum is partitioned
into $2^n$ {\em patterns}, corresponding to all possible subsets
of active APs. The problem is to optimize
the widths of spectrum segments, associated with the different patterns,
along with the user association under each pattern.
This has been shown to provide significant performance improvement
in throughput enhancement and delay reduction
\cite{zhuang2014traffic,zhuang2016scalable,kuang2014joint,kuang2016energy}.

The original convex problem formulation in \cite{zhuang2014traffic}
is not useful for large networks because the number of patterns
grows as $O(2^n)$.
In prior work~\cite{zhuang2016scalable}, we have reduced the number of variables
to $O(n)$ by recognizing that each link rate depends only on {\em local} patterns
of active APs. The problem can then be redefined over sets of overlapping
{\em interference neighborhoods}, associated with those local patterns.
Each AP has its own interference neighborhood, which
captures the interference from nearby APs. The challenge with
the approach in~\cite{zhuang2016scalable} is to ensure
that the spectrum assigned to each particular AP
is consistent across the neighborhoods to which it belongs.
To accomplish that, the spectrum is discretized and a coloring algorithm is
proposed to ensure that the local patterns of active APs are globally consistent.

Here we take a different approach to addressing the scalability problem.
This is based on the fact that the solution is {\em sparse},
meaning that at most $k+1$ out of the $2^n$ possible patterns
appear in the optimal allocation for a general network utility function.
Hence we reformulate the problem by dividing the spectrum into $k+1$ segments,
rather than $2^n$, and attempt to identify the pattern
that should be associated with each segment. This effectively
reverses the approach in~\cite{zhuang2016scalable}, which attempts
to assign a segment of spectrum to each pattern.
In this reformulation, we initially assume that any combination of patterns
can be assigned to each of the $k+1$ segments.
This problem is a convex relaxation of our original problem.
The one-to-one mapping of spectrum segments to patterns is then enforced
with an $\ell_0$ (cardinality) constraint.
An algorithm for finding an approximate solution to this problem is presented
based on a reweighted $\ell_1$ approximation of this constraint \cite{candes2008enhancing}.

The approach to scalability presented here has the advantage
of eliminating the combinatorial coloring problem that arises in~\cite{zhuang2016scalable}.
Instead, a new $\ell_0$ constraint is introduced. Although this does not
simplify the original problem, it helps in finding an approximate solution,
since the reweighted $\ell_1$ approximations for the $\ell_0$ constraints
are known to perform well.
Numerical results indicate that this method generally gives better performance
for a fixed computational complexity than the method in~\cite{zhuang2016scalable}.


\section{System Model}
\label{sec:SysMod}

\vspace{-5pt}
\begin{figure}
\centering
\includegraphics[width=0.9\columnwidth]{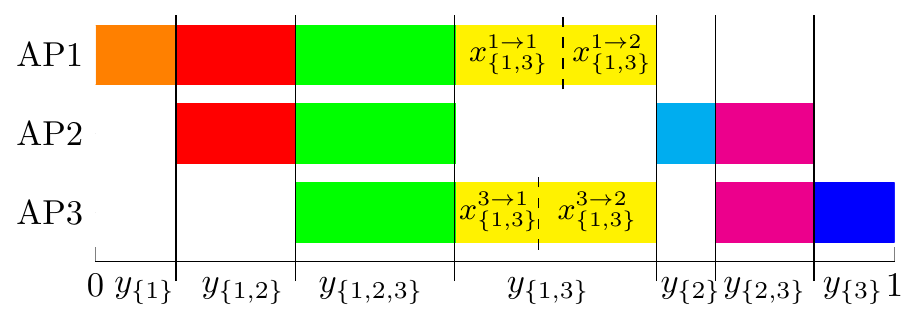}\vspace{-5pt}
\caption{An example off spectrum allocation among 3 APs.}
\label{fig:Mod}
\end{figure}

As in~\cite{zhuang2016scalable}, we consider a network containing
the set of APs $\NN=\{1,\cdots,n\}$ and the set of UEs $\KK=\{1,\cdots,k\}$.
Each ``UE'' is actually associated with a particular location, and could refer
to a group of nearby mobiles.
The $n$ APs share $W$ Hz of spectrum. For convenience, we normalize $W=1$.
Each AP can transmit on any part(s) of the spectrum, which is homogeneous
(has the same distance attenuation). APs sharing the same spectrum
interfere at UEs within range of both transmitters. All transmissions are assumed
to be omni-directional, although a similar set of problems can be reformulated
with directional transmissions.

We define a subset of active (transmitting) APs
$A\subset\NN$ as a {\em pattern}.
There are $2^n$ patterns, and a particular allocation of spectrum maps
each pattern to a slice of spectrum.
We denote an allocation as $\{y_A\}_{A\subset \NN}$,
where $y_A$ denotes the amount of bandwidth assigned to pattern $A$.
An example with three APs is depicted in Fig.~\ref{fig:Mod}.
AP 1 \EW{owns ${\{1\}}$ exclusively}; shares ${\{1,2\}}$ with AP 2;
shares ${\{1,3\}}$ with AP 3; and shares ${\{1,2,3\}}$ with both AP 2 and AP 3.

User association is determined by how the spectrum assigned to a particular AP
is allocated to different UEs.  \EW{Specifically,} denote the fraction of total bandwidth
used by AP $i$ to serve UE $j$ under pattern $A$ as $x_A^{i\to j}$,
for $i\in A$.
UE $j$ is then assigned to AP $i$ if $x_A^{i\to j}>0$ for some $A \subset \mathcal{N}$.
Since the total bandwidth assigned to pattern $A$ is $y_A$, we have
\begin{align}
\label{eq:sum_x=y}
\sum_{j\in\KK}x^{i\to j}_A \leq y_A,~\forall A\subset\NN,\;i\in A.
\end{align}
The total bandwidth allocated to all patterns
is then
\begin{align}
\label{eq:bandwidth}
\sum_{A\subset \NN}y_A=1.
\end{align}

Let $s_A^{i\to j}$ denote the spectral efficiency of the link from AP $i$ to UE $j$ under pattern $A$.
This is measured over a {\em slow timescale}, and is therefore an average over short-term fading.
The value of $s_A^{i\to j}$ is therefore determined by the distance between AP $i$ and UE $j$,
shadowing, and similar long-term characteristics of the interference links from APs in $A$
to UE location $j$.
For concreteness
we assume
\begin{align} \label{eq:SE}
  s^{i\rightarrow j}_A
  =
  \frac{W}{\tau}\mathbbm{1}_{\{i\in A\}}
  \log_2\left(1+\frac{p_ig^{i\to j}}{\sum\limits_{i'\in A\setminus\{i\}}p_{i'}g^{i'\to j}+n_j}\right) 
\end{align}
where $\mathbbm{1}_{\{i\in A\}}=1$ if $i\in A$ and $0$ otherwise,
$p_i$ is the transmit power spectral density (PSD) at AP $i$, $g^{i\to j}$ is the power gain of link
{$i\to j$}, and $n_j$ is the noise PSD
at UE $j$. We assume fixed, flat transmit PSDs over the slow timescale considered.
The factor $W/\tau$, where $\tau$ is the average packet length (bits), gives
the units in packets/sec. The link gain $g^{i\to j}$ includes pathloss and shadowing effects.
Clearly, $s_A^{i\to j}=0$ if $i\not\in A$, i.e., AP $i$ does not transmit on pattern $A$.
In practice, the spectral efficiencies can be measured as time-averaged channel gains.
The total rate received by UE $j$ is therefore
\begin{align}
\label{eq:rate}
r^j=\sum_{A\subset \NN}\sum_{i\in A}s_A^{i\to j}x^{i\to j}_A,~\forall j\in\KK.
\end{align}

\section{Problem Formulation with Global Patterns}
\label{sec:GlobalForm}
The problem is to maximize a network utility function over the spectrum allocation and user association
designated by
$\bx=\big(x^{i\to j}_A\big)_{A\subset \NN, i\in A, j\in\KK}$\EW{, $\by = (y_A)_{A\subset \NN}$:}
\begin{varsubequations}{P0}
\label{eq:Opt}
\begin{align}
  \maximize_{\br,\;\bx,\;\by}~
  & u(r^1,\cdots,r^k) \label{eq:Obj-Opt}\\
  \subjectto~&
   x^{i\to j}_A\geq 0,
  \quad \forall A\subset \NN,\;\forall i\in A,~\forall j\in \KK \label{eq:Con4-Opt}
\end{align}
\end{varsubequations}\noindent
and constraints \eqref{eq:rate}, \eqref{eq:sum_x=y}, and \eqref{eq:bandwidth},
where the network utility function $u$ depends on the service rates to all UEs.
The optimization problem is convex if $u$
is concave in $\br=[r^1,\cdots,r^k]$. \sout{and element-wise non-decreasing.}
As in~\cite{zhuang2016scalable}, we will take $u$ to be the average packet delay, given by
\begin{align}\label{eq:AvgDelay}
  u(r^1,\cdots,r^k)=-\sum_j\frac{\lambda^j}{(r^j-\lambda^j)^+}
\end{align}\noindent
where $\frac{1}{(x)^+}$ equals $\frac{1}{x}$ if $x>0$ and $+\infty$ otherwise, and $\lambda_j$ is the Poisson packet arrival rate for UE $j$.
This assumes exponential packet lengths and backlogged interference~\cite{zhuang2014traffic}.

Solving~\ref{eq:Opt} becomes prohibitively expensive 
as the network size grows, due to the inherit complexity from the $2^n$ global patterns.
However, a key property of solution(s)
to \eqref{eq:Opt} is that at least one is {\em sparse}, i.e., contains at most $k+1$ patterns.
\begin{proposition}{(\cite{zhuang2015energy})}
\label{thm:Spec}
\ref{eq:Opt} has a solution that divides
the spectrum into at most $k+1$ segments, i.e.,
\begin{align} \label{eq:ThmSpec}
  \left|\{A\subset \NN~|~y_A>0\}\right|\leq k+1.
\end{align}\nonumber
Furthermore, if $u$ is element-wise nondecreasing in $\br$,
$k+1$ in \eqref{eq:ThmSpec} is reduced to $k$.
\end{proposition}
Determining which $k+1$ active patterns appear in a solution
is then the key challenge in solving~\ref{eq:Opt}.

\section{Reformulation With Sparsity Constraints}
\label{sec:MainRes}
\subsection{Local Neighborhoods}
\label{subsec:local}
We first reduce the number of variables in \eqref{eq:Opt} from $O(2^n)$ to $O(n)$
by reformulating \eqref{eq:Opt} over {\em interference neighborhoods}
based on local patterns~\cite{zhuang2016scalable}. Due to pathloss, we assume interference vanishes
beyond a certain distance. Let $L\subset \NN\times \KK$ denote the set of links with nonzero gains.
Hence each UE only receives power from APs within its neighborhood:
\vspace{-0.2cm}\begin{align}
\label{eq:SetA}
\mathcal{A}_j=\{i|(i\to j)\in L\}.
\end{align}
From the AP side, each AP can only transmit to a collection of UEs in a AP neighborhood with positive rates:
\vspace{-0.1cm}\begin{align}
\label{eq:SetU}
\mathcal{U}_i=\{j|(i\to j)\in L\}.
\end{align}\vspace{-1pt}
We define the interference neighborhood for each AP $i$ as:
\begin{align}
\label{eq:SetN}
\mathcal{N}_i = \cup_{j\in\mathcal{U}_i}\mathcal{A}_j,
\end{align}
i.e., $\mathcal{N}_i$ includes AP $i$ and all APs that interfere with it.

\begin{figure}
\centering
\includegraphics[width=.6\columnwidth]{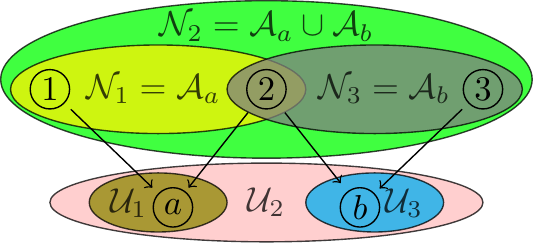} 
\caption{Neighborhoods in the case of three APs and two UEs.}
\label{fig:LocalPattern}
\end{figure}

An example with three APs (denoted by $\{1,2,3\}$) and 2 UEs (denoted by $\{a,b\}$) is shown in Fig~\ref{fig:LocalPattern}. The set of nonzero links are $L=\{1\to a,\;2\to a,\;2\to b,\;3\to b\}$.
The AP neighborhoods are $\mathcal{U}_1=\{a\}$, $\mathcal{U}_2=\{a,b\}$, and $\mathcal{U}_3=\{b\}$; and the UE neighborhoods are $\mathcal{A}_a=\{1,2\}$ and $\mathcal{A}_b=\{2,3\}$. AP 1's interference neighborhood is $\mathcal{N}_1=\{1,2\}$, as AP 2 interferes with it at UE $a$. AP 3's interference neighborhood is $\mathcal{N}_3=\{2,3\}$, as AP 2 interferes with it at UE $b$. AP 2's interference neighborhood is $\mathcal{N}_2=\{1,2,3\}$, since AP 1 and AP 3 interfere with AP 2 at UE $a$ and UE $b$, respectively.

The spectral efficiency of link $i\to j$ only depends on the local patterns in $\mathcal{A}_j$, according to the definition of UE neighborhoods in~\eqref{eq:SetA}. Therefore, for any link $i\to j$, \EW{the spectral efficiency under }a global pattern $A$ is equivalent to its intersection with UE $j$'s neighborhood $\mathcal{A}_j$:
\vspace{-5pt}\begin{align}
\label{eq:LocalSE}
s^{i\to j}_A=s^{i\to j}_{A\cap \mathcal{A}_j} ~\forall j\in\KK,\; \forall i\in \mathcal{A}_j,\; A\subset\NN.
\end{align}

\vspace{-5pt}We next express each rate in terms of local patterns.
We define bandwidth allocation variables within a local interference neighborhood $\NN_i$ as:
\vspace{-5pt}\begin{align}
\label{eq:LocalGlobal}
z^{i\to j}_B=\sum_{A\subset\NN:A\cap\NN_i=B}x^{i\to j}_A,~i\in\NN,\;j\in\mathcal{U}_i,\;\forall B\subset\NN_i.\vspace{-5pt}
\end{align}
\sout{If we treat an interference cluster $\NN_i$ as a partition of all the global patterns, many global patterns are equivalent.}
Because all global patterns sharing the same overlap with $\NN_i$ contribute to the same local pattern of AP $i$, $z^{i\to j}_B$ represents the bandwidth allocated to link $i\to j$ under local pattern $B$.
Hence from AP $i$'s perspective, the bandwidth assigned to a local pattern $B$ must be the sum of the bandwidths assigned to all global patterns containing $B$.
The service rate of link $i\to j$, as defined in~\eqref{eq:rate}, can then be calculated
as \cite{zhuang2016scalable}:
\begin{equation}
r^j
=\sum_{i\in \mathcal{A}_j}\sum_{B\subset\NN_i}s^{i\to j}_{B\cap\mathcal{A}_j}z^{i\to j}_B 
\label{eq:LocalRate}
\end{equation}%
where we have used~\eqref{eq:LocalSE} and~\eqref{eq:LocalGlobal},
and the fact that only APs in $\mathcal{A}_j$ transmit to UE $j$ with positive rate.

\subsection{Sparse Optimization}
\EW{Motivated by Proposition \ref{thm:Spec},
we reformulate \eqref{eq:Opt} by dividing the spectrum into $k+1$ segments,
and seek to assign a single pattern to each segment.} Proposition \ref{thm:Spec}
implies that by optimizing this assignment we obtain a solution to \ref{eq:Opt}. \sout{The number of local patterns in each local interference cluster is bounded despite the size of the network. The general idea is provided here.}
Let $h_l$ denote the bandwidth of the $l$th segment (to be optimized).
We now associate a set of variables $z$ (for local neighborhoods)
and $y$ {\em for each segment} $l$.
We can therefore rewrite the rate in~\eqref{eq:LocalRate} as
\begin{align}
\label{eq:Con1}
r^j=\sum_{l=1}^{k+1}\sum_{i\in\mathcal{A}_j}\sum_{B\subset\NN_i} s^{i\to j}_{B\cap\mathcal{A}_j}\;z^{i\to j}_{B,l},~\forall j\in \KK
\end{align}
where the first sum is over the $k+1$ spectrum segments.
Apart from the addition of segment index, the $z$ variable is the same local bandwidth allocation variable defined in~\eqref{eq:LocalGlobal}.
The amount of bandwidth assigned to local pattern $B$ in AP $i$'s interference neighborhood
$\NN_i$ within segment $l$ is
\begin{align}
\label{eq:Con2}
y^i_{B,l}=\sum_{j\in\mathcal{U}_i}z^{i\to j}_{B,l}, ~\forall i\in\NN,\;\forall B\subset\NN_i .
\end{align}
The total amount of spectrum assigned to $\NN_i$ satisfies
\begin{align}
\label{eq:Con5}
\sum_{B\subset \NN_i}y^i_{B,l}\leq h_l,~\forall i.
\end{align}

We also introduce the following {\em consistency constraint} to ensure that
the amount of spectrum allocated to an AP is consistent across  any
two neighborhoods $\NN_i$ and $\NN_m$ that contain it~\cite{zhuang2016scalable}.
That is, for any
nonempty $C\subset\NN_i\cap\NN_m$, 
\begin{align}
\label{eq:Con3}
\sum_{B\subset \NN_i:B\cap\NN_m=C}y^i_{B,l}=\sum_{B\subset \NN_m:B\cap\NN_i=C}y^m_{B,l}.
\end{align}
See, for example, $\NN_1$ and $\NN_3$ in Fig.~\ref{fig:LocalPattern}. In interference cluster $\NN_1$, AP 2 transmits under pattern $\{2\}$ and $\{1,2\}$; in interference cluster $\NN_3$, AP 2 transmits under pattern $\{2\}$ and $\{2,3\}$. The total bandwidth used by AP 2 must be consistent across
neighborhoods so that
$y^1_{\{2\},l}+y^1_{\{1,2\},l}=y^3_{\{2\},l}+y^3_{\{2,3\},l}$, where $\{2\}$ is the overlapping pattern.
This example can be extended to a {\em set} of APs, which are members of two
interference neighborhoods, giving~\eqref{eq:Con3}.

To ensure a one-to-one mapping of patterns to the $k+1$ segments,
we add the $\ell_0$-norm constraint
\begin{align}
\label{eq:Con6}
\sum_{B\subset\NN_i}|y^i_{B,l}|_0\leq1,&~\forall i\in\NN,\; l=1,\cdots,k+1
\end{align}
where $|x|_0=1$ if $x\neq 0$, and $|x|_0=0$ if $x=0$.
That is, we constrain each AP to use at most one active pattern in each segment.

We can now reformulate \ref{eq:Opt} in terms of the local interference variables $\bz$
across the $k+1$ spectrum segments:
\begin{varsubequations}{P1}
\label{eq:Opt3}
\begin{align}
&\maximize_{\br,\;\by,\;\bz} \;
u(r^1,\cdots,r^k) & \label{eq:Obj-Opt3}\\
& \subjectto \;
  z^{i\to j}_{B,l}\geq 0,~~
  \forall (i\to j)\in L,\;
  \forall B\subset\NN_i,\;
  \label{eq:Con9-Opt3}\vspace{-2cm}\\
  &
  \qquad \sum_{l=1,\cdots,k+1}h_l=1, ~~~h_l\geq0,
   \label{eq:Con4-Opt3}\vspace{-0.2cm}
\end{align}
\end{varsubequations}\noindent
and constraints \eqref{eq:Con1}-\eqref{eq:Con6} for each $l=1,\cdots,k+1$.
\begin{theorem}
\label{thm:main}
\ref{eq:Opt} and~\ref{eq:Opt3} are equivalent (have the same solutions)
given the local neighborhood definitions~\eqref{eq:SetA}-\eqref{eq:SetN}.
\end{theorem}

The proof is omitted due to limited space.
Hence a solution to~\ref{eq:Opt3} always satisfies Proposition~\ref{thm:Spec}.
This reformulation leads to a computationally efficient approximation algorithm.

\section{Iterative 
Approximation}
\label{sec:l1}
The number of variables is reduced from $O(2^n)$ in~\ref{eq:Opt} to $O(nk)$ in~\ref{eq:Opt3}. However, the $\ell_0$ norm constraint makes the problem non-convex and difficult to solve.
We apply the reweighted $\ell_1$ approach, described in~\cite{candes2008enhancing},
to approximate the $\ell_0$ constraint.
Specifically, the $\ell_0$ norm is approximated by the weighted $\ell_1$ norm, $\sum_i w_i|x_i|$,
where $w_i$ is iteratively adapted, and taken to be inversely proportional
to $|x_i|$ computed at the preceding iteration.
This has the effect of suppressing small nonzero entries with large
weights.\footnote{This algorithm has also been used in~\cite{zhuang2015energy}
to solve an AP activation problem.}

The $\ell_1$ approximation cannot be directly applied to~\ref{eq:Opt3}
since the $\ell_0$ norm constraints are coupled through~\eqref{eq:Con6} and
\eqref{eq:Con4-Opt3}.
Hence we present an iterative algorithm based on the $\ell_1$ reweighted heuristic,
where the weights depend on both the $\ell_1$ norm and the bandwidth of each segment $h_l$.
In each iteration of the algorithm, shown in Algorithm~\ref{alg:l1},
we solve~\ref{eq:Opt3}, but with the $\ell_0$ norm constraint~\eqref{eq:Con6}
replaced by the weighted sum
 \vspace{-5pt}
\begin{equation}
\sum_{B\subset\NN_i}w^i_{B,l}y^i_{B,l}\leq1 .
\label{eq:l1approx}\vspace{-3pt}
\end{equation}
In Algorithm~\ref{alg:l1}, we refer to this reweighted $\ell_1$ version of \ref{eq:Opt3}
as P2.


Algorithm \ref{alg:l1} shows a random initialization of the weights to introduce
asymmetry in the first iteration. Otherwise, e.g., if all $w^i_{B,l}$'s are initialized
to a constant,
the solution stays symmetric over all segments,
because the constraints and objectives are identical for each segment.
A symmetric solution, i.e., $y^i_{B,1}=y^i_{B,2}=,\cdots,=y^i_{B,k},~\forall i\in \NN,~B\subset\NN_i$, \sout{is obviously not optimal for an arbitrary network topology}\EW{is generally not a solution to the original problem}. In each iteration, we solve P2 with the current weights $\bw$ to obtain the current $\bx,~\by,~\bz$ and $\bh=[h_1,\cdots,h_k]$. Then the weights are updated
as shown.
The iterations terminate when the variables converge
or the maximum number of iterations $t_{max}$ is reached.

The weight update in Algorithm~\ref{alg:l1} is obtained by approximating the $\ell_0$ norm
with $\log(x+\epsilon)$ for small $\epsilon$ (see~\cite{candes2008enhancing} and the references therein).
Adding $\epsilon$ in the denominator allows small components to be propagated to the next iteration.
Here the goal is to obtain a {\em single} nonzero $y^i_{B,l}=h_l,~B\in\NN_i$, so we take
$\epsilon=\alpha h_l$. This is different from the fixed $\epsilon$
proposed in~\cite{candes2008enhancing} in that $\alpha h_l$ changes
with $h_l$ in each iteration.
That is, Algorithm~\ref{alg:l1} simultaneously searches
for the optimal reuse pattern and its associated bandwidth.

It is possible that Algorithm~\ref{alg:l1} produces multiple reuse patterns for a segment at termination.
In those cases AP $i$ chooses the dominant pattern
$B^i_l= \arg\max_{B\in\NN_i} y^i_{B,l}$. The global reuse pattern assigned to segment $l$
is then given by $A^*_l=\cup_{i:i\in B^i_l}\{i\}$. 
The spectral efficiencies for each segment are determined by
the corresponding global reuse pattern (set of interfering APs):
$\bar{s}^{i\to j}_l=s^{i\to j}_{A^*_l\cap \mathcal{A}_j}$.
Given the assignment of patterns to each spectrum segment,
the widths of the segments, $\bh=(h_l)_{l=1,\cdots,k+1}$ along with
the assignment of spectrum to mobiles, $\bar{\bx}= (\bar{x}^{i\to j}_l)_{j\in \KK,i\in\mathcal{A}_j,l=1,\cdots,k+1}$,
can then be re-optimized by solving the convex problem:
\begin{varsubequations}{P3}\vspace{-3pt}
\label{eq:Opt7}
\begin{align}
  \maximize_{\bar{\bx},\;\bh} \;
  & u(r_1,\cdots,r_k) \label{eq:Obj-Opt7}\\\vspace{-3pt}
   \subjectto \;
  & r_j = \sum_{l=1}^k\sum_{i\in \mathcal{A}_j} \bar{s}^{i\to j}_l \bar{x}_l^{i\to j},
  \quad \forall j\in\KK \label{eq:Con1-Opt7}\\
  &  \sum_{j=1}^k \bar{x}^{i\to j}_l\leq h_l,~\forall j\in \KK, i\in \mathcal{A}_j
  \label{eq:Con2-Opt7}
\end{align}\vspace{-5pt}
\end{varsubequations}
$\bar{x}^{i\to j}_l \geq 0$ and \eqref{eq:Con4-Opt3} for $l=1,\cdots,k+1$.

\begin{algorithm}
\caption{Iterative re-weighted $\ell_1$ approximation.}
\label{alg:l1}
\begin{algorithmic}[]
\INPUT {$(s^{i\to j}_C)_{j\in \KK, i\in \mathcal{A}_j, C\subset \mathcal{A}_j}$, and $(\lambda_j)_{j\in \Kcal}$.}
\OUTPUT{The widths of $k+1$ segments  $(h_l)_{l=1,\cdots,k+1}$, the $k+1$ active patterns $(A^*_l)_{l=1,\cdots,k+1}$, and the spectrum allocated to link $i\to j$ on segment $l$, $(\bar{x}^{i\to j}_l)_{j\in \KK,i\in\mathcal{A}_j,l=1,\cdots,k+1}$}
\Init{Choose $w_{B,l}^i$'s randomly in (0,1).\\
Set iteration counter $t=0$.}
\While{
Variables have not converged and $t<t_{max}$}
    \State 1. Solve P2
    with the current weights $\bw$.
    \State 2. Update $w^i_{B,l}=\frac{1}{y^i_{B,l}+\alpha h_l}$.
    \State 3. $t=t+1$.
\EndWhile
\Post {Determine the reuse patterns $\{A^*_l \}$
and spectral efficiencies $\{ \bar{s}^{i\to j}_l\}$ across segments;
solve P3.}
\end{algorithmic}
\end{algorithm}\vspace{-0.1cm}

\section{Numerical Results}
\label{sec:NumRes}
The simulation results assume one macro AP is located at the center of the area,
and the remaining $n-1$ pico APs are randomly dropped around it.
The $k$ UEs are placed on a rectangular lattice.
Link gains include both pathloss and shadowing.
Additional simulation parameters are shown
in the footnote.\footnote{The pathloss exponent is 3, standard deviation
of shadow fading is 3, macro-transmit PSD is 5 $\mu$W/Hz,
pico transmit PSD is 1 $\mu$W/Hz,
noise PSD is $10^{-7}$ $\mu$W/Hz,
total bandwidth is 20 MHz, and
average packet length is 1 Mb.}

\subsection{Small Network}
\label{subsec:small}
We first compare solutions to~\ref{eq:Opt} and~\ref{eq:Opt3} for a small network with $n=10$ and $k=32$. 
Since the number of variables is relatively small, we solve both versions of~\ref{eq:Opt} with and without the local neighborhood approximation using a standard convex optimization solver. The local neighborhoods are constructed by including the strongest four APs for each UE.
The solution to~\ref{eq:Opt3} is obtained using Algorithm~\ref{alg:l1}.
%
We compare those with full spectrum reuse where each user is assigned
to the AP with the strongest signal (maxRSRP association), and also
the optimal orthogonal allocation,\footnote{Both spectrum allocation
and user association are optimized assuming each AP exclusively occupies a fraction of the spectrum.} i.e., only $\{y_{\{i\}}\}_{i\in\NN}$ are active.

\begin{figure}
\centering
\includegraphics[width=0.9\columnwidth, height=65mm]{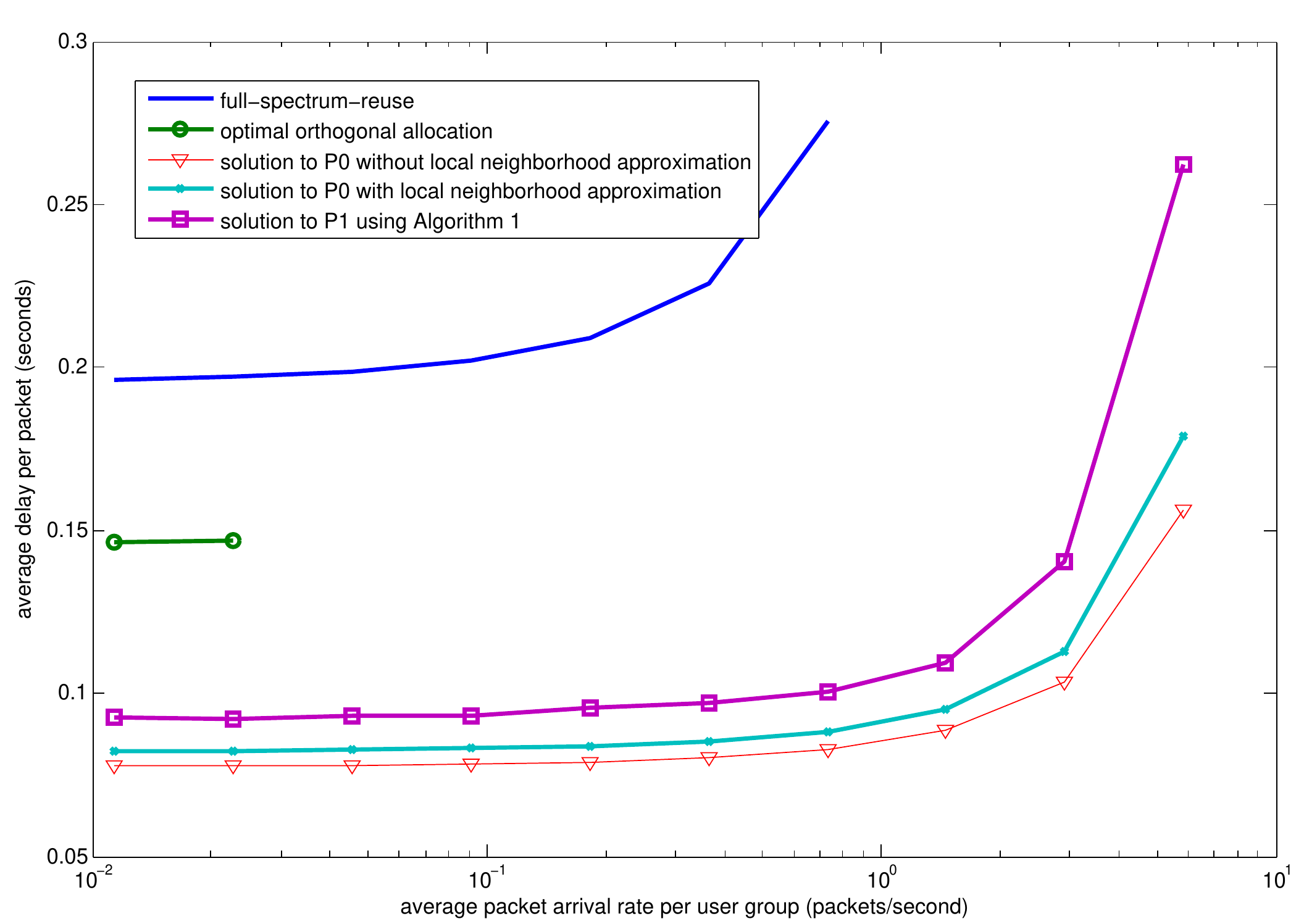}\vspace{-0.1cm}
\caption{Delay versus traffic arrival rate curves for a small network with $n=10$ and $k=23$.}
\label{fig:delay-N10K23}
\end{figure}

Fig. ~\ref{fig:delay-N10K23} shows delay versus traffic arrival rate for all schemes.
The end of each curve represents the maximum arrival rate the scheme can support.
The curves obtained by solving~\ref{eq:Opt} with and without local neighborhood approximation are very close, which indicates considering the four strongest interferers is enough in such a small network.
The solution to~\ref{eq:Opt3} incurs slightly larger delay than the solution to~\ref{eq:Opt}.
The jointly optimized spectrum allocations and user associations achieve substantial delay reduction as well as eight times throughput compared to full frequency reuse with maxRSRP association.

\begin{figure} 
\centering \vspace{-0.2cm}
\includegraphics[width=0.65\columnwidth, height=60mm]{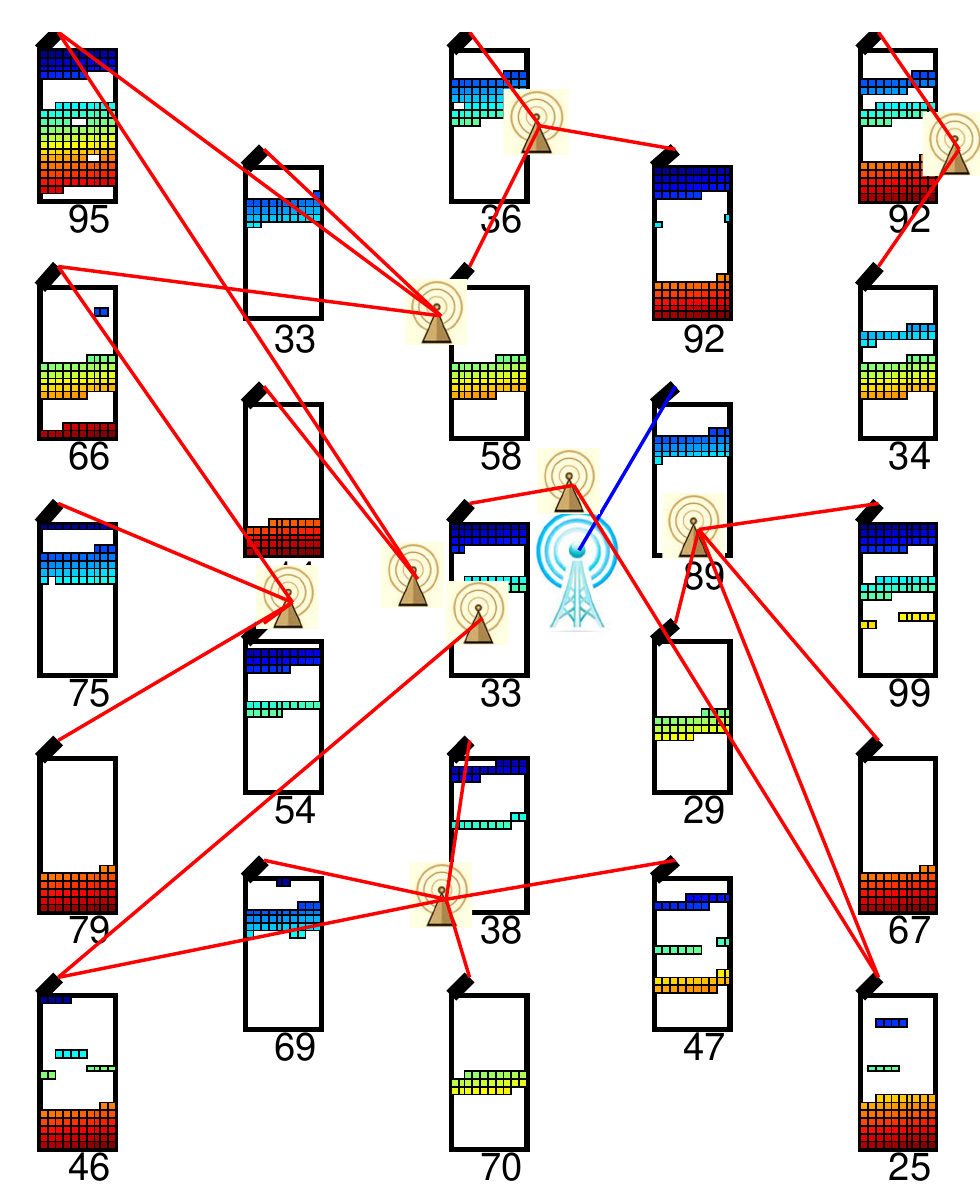}\vspace{-0.1cm}
\caption{Example spectrum allocations and user associations from \eqref{eq:Opt3}.}
\setlength{\belowcaptionskip}{-10pt}\vspace{-0.5cm}
\label{fig:Sol3}
\end{figure}

The optimized spectrum allocation and user associations obtained
solving~\ref{eq:Opt3} are depicted in
Fig.~\ref{fig:Sol3}.
The macro and pico APs are represented by the bigger and smaller towers;
each handset represents a UE. Each solid line shows a connection between
the corresponding AP and UE. The grid for each UE shows
the spectrum used to serve that UE.
The traffic arrival rate (in (0, 100)) for each UE is shown under each grid.
The user association in Fig.~\ref{fig:Sol3} is close but not identical to that
obtained by solving~\ref{eq:Opt} (not shown)
due to the approximation in Algorithm~\ref{alg:l1},
which explains the performance difference
shown in Fig.~\ref{fig:delay-N10K23}.



\subsection{Large Network}
\begin{figure}
\centering
\includegraphics[width=0.96\columnwidth,height=60mm]{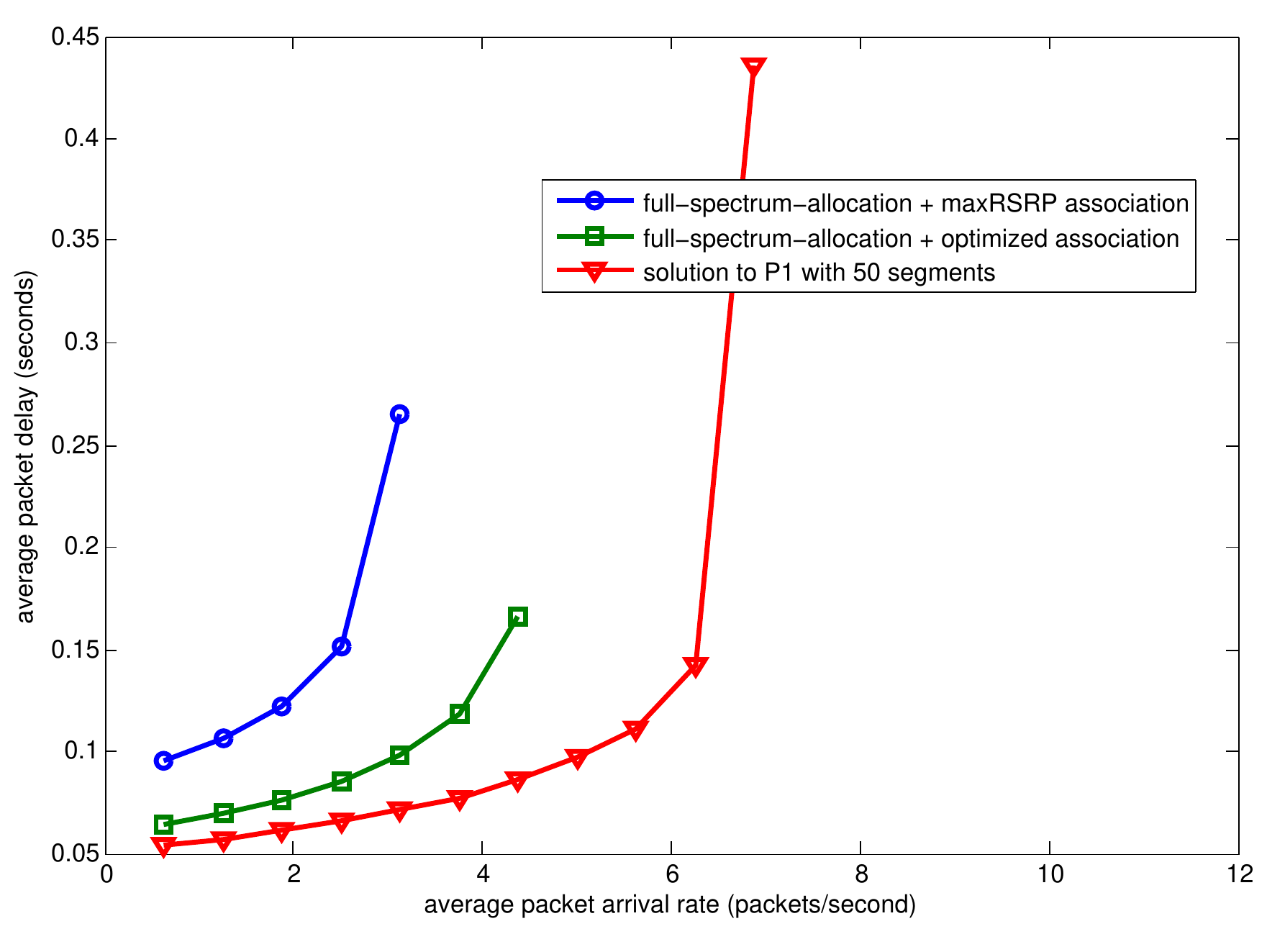}\vspace{-0.2cm}
\caption{Delay versus traffic arrival rate curves for a large network with $n=100$ and $k=195$.}\vspace{-0.4cm}
\label{fig:delay-N100K195}
\end{figure}

Fig. \ref{fig:delay-N100K195} compares the performance of different allocation schemes
in a large network with $n=100$ APs and $k=200$ UEs.
To facilitate the simulations, we reduce the size of each UE neighborhood from four to three,
i.e., each UE can only be served by the three strongest APs.
We compare full-spectrum-reuse with maxRSRP association, full-spectrum-reuse with optimized associations and the solution to~\ref{eq:Opt3} with 50 segments.
The solution to~\ref{eq:Opt3} achieves 1.5 times the throughput
of the full-spectrum-reuse with optimized association.
The solutions to~\ref{eq:Opt3} use no more than 23 active patterns ($<50$ available segments).
The throughput gain achieved by the solution to~\ref{eq:Opt3} in this large network
is smaller than that in Section~\ref{subsec:small}.
This is mainly because we only consider the three strongest interferers,
which compromises the benefits from interference management.

\section{Conclusion}
\label{sec:Con}
An approach to joint allocation of spectrum with user association has been
presented which exploits the sparsity of the optimal solution.
The proposed algorithm has been observed to achieve near-optimal
performance for small to medium-size networks for which the optimal solution can be computed,
and provides substantial gains relative to full frequency reuse.
Although not considered here, the formulation can be extended
to accommodate spatial selectivity and different power levels.
The performance-complexity tradeoff for such extensions is left for future work.

\bibliographystyle{ieeetr}


\begin{thebibliography}{1}

\bibitem{yu2013multicell}
W.~Yu, T.~Kwon, and C.~Shin, ``Multicell coordination via joint scheduling,
  beamforming, and power spectrum adaptation,'' {\em {IEEE} Trans. Wireless
  Commun.}, vol.~12, no.~7, pp.~1--14, 2013.

\bibitem{wang2013joint}
F.~Wang, L.~Song, Z.~Han, Q.~Zhao, and X.~Wang, ``Joint scheduling and resource
  allocation for device-to-device underlay communication,'' in {\em Proc.\
  Conf.\ Wireless \ Comm. and Networking}, pp.~134--139, IEEE, 2013.

\bibitem{HonLuo13TransSP}
M.~Hong and Z.-Q. Luo, ``Distributed linear precoder optimization and base
  station selection for an uplink heterogeneous network,'' {\em {IEEE} Trans.
  Signal Process.}, vol.~61, pp.~3214--3228, June 2013.

\bibitem{zhuang2014traffic}
B.~Zhuang, D.~Guo, and M.~L. Honig, ``Traffic-driven spectrum allocation in
  heterogeneous networks,'' {\em {IEEE} J. Sel. Areas Commun.}, vol.~PP,
  no.~99, pp.~1--1, 2015.

\bibitem{kuang2014joint}
Q.~Kuang, ``Joint user association and reuse pattern selection in heterogeneous
  networks,'' in {\em 2014 11th International Symposium on Wireless
  Communications Systems (ISWCS)}, pp.~401--405, IEEE, 2014.

\bibitem{zhuang2016scalable}
B.~Zhuang, D.~Guo, E.~Wei, and M.~L. Honig, ``Scalable spectrum allocation and
  user association in networks with many small cells,''
 {\em https://arxiv.org/abs/1701.03247,} submitted to {\em IEEE Trans. Commun.}.

\bibitem{kuang2016energy}
Q.~Kuang and W.~Utschick, ``Energy management in heterogeneous networks with
  cell activation, user association, and interference coordination,'' {\em IEEE
  Trans. Wireless Commun.}, vol.~15, June 2016.

\bibitem{candes2008enhancing}
E.~J. Candes, M.~B. Wakin, and S.~P. Boyd, ``Enhancing sparsity by reweighted
  $l_1$ minimization,'' {\em Journal of Fourier analysis and applications},
  vol.~14, no.~5-6, pp.~877--905, 2008.

\bibitem{zhuang2015energy}
B.~Zhuang, D.~Guo, and M.~L. Honig, ``Energy-efficient cell activation, user
  association, and spectrum allocation in heterogeneous networks,'' {\em IEEE
  J. Sel. Areas Commun.}, vol.~PP, no.~99, pp.~1--1,
  2016.

\end{thebibliography}
\end{document}